\documentclass[reqno,11pt]{amsart}
\usepackage{amsmath, latexsym, amsfonts, amssymb}
\usepackage{mathrsfs,enumerate}
\usepackage{graphics,epsfig,color}

\numberwithin{equation}{section}

\newtheorem{theorem}{Theorem}[section]
\newtheorem{conjecture}[theorem]{Conjecture}

\newtheorem*{Conditions}{Crossing Conditions}

\newcommand{\cP}{{\ensuremath{\mathcal P}} }

\newcommand{\cN}{{\ensuremath{\mathcal N}} }
\newcommand{\cO}{{\ensuremath{\mathcal O}} }

\newcommand{\cM}{{\ensuremath{\mathcal M}} }

\newcommand{\cQ}{{\ensuremath{\mathcal Q}} }

\newcommand{\bbE}{{\ensuremath{\mathbb E}} }

\newcommand{\bbN}{{\ensuremath{\mathbb N}} }

\newcommand{\bbP}{{\ensuremath{\mathbb P}} }
\newcommand{\bbQ}{{\ensuremath{\mathbb Q}} }
\newcommand{\bbR}{{\ensuremath{\mathbb R}} }

\newcommand{\bbZ}{{\ensuremath{\mathbb Z}} }

\newfont{\indic}{bbmss12}

\newcommand{\q}{{\ensuremath{\mathbf q}} }
\newcommand{\p}{{\ensuremath{\mathbf p}} }
\newcommand{\z}{{\ensuremath{\mathbf z}} }
\newcommand{\x}{{\ensuremath{\mathbf x}} }
\newcommand{\e}{{\ensuremath{\mathbf e}} }

\begin{document}
\title[The Mirrors Model : Macroscopic Diffusion Without Noise or Chaos ]{The Mirrors Model : Macroscopic Diffusion Without Noise or Chaos }
\author[Y.Chiffaudel]{Yann Chiffaudel}
 \address{Laboratoire de Probabilit\'es
  et Mod\`eles Al\'eatoires (CNRS UMR 7599), Universit\'e Paris Diderot,
UFR de Math\'ematiques, b\^atiment Sophie Germain,
5 rue Thomas Mann,
75205 Paris CEDEX 13
France}
\email{}
\author[R.\ Lefevere]{Rapha\"el Lefevere}
 \address{Laboratoire de Probabilit\'es
  et Mod\`eles Al\'eatoires (CNRS UMR 7599), Universit\'e Paris Diderot,
UFR de Math\'ematiques, b\^atiment Sophie Germain,
5 rue Thomas Mann,
75205 Paris CEDEX 13
France}
\email{lefevere\@@math.univ-paris-diderot.fr}

 

\maketitle

\begin{abstract}
Before stating our main result, we first clarify through classical examples the status of the laws of macroscopic physics as laws of large numbers.
We next consider the mirrors model in a finite $d$-dimensional domain and connected to particles reservoirs at fixed chemical potentials. The dynamics is purely deterministic and non-ergodic but takes place in a random environment. We study  the macroscopic current  of particles in the stationary regime. We show first that when the size of the system goes to infinity, the behaviour of the stationary current of particles is governed by the proportion of orbits crossing the system.  This allows to formulate a necessary and sufficient  condition on the distribution of the set of orbits that ensures the validity of Fick's law. Using this approach, we show  that Fick's law relating the stationary macroscopic current of particles to the concentration difference holds in three dimensions and above.   The negative correlations between crossing orbits play a key role in the argument.
 \end{abstract}
\section{Macroscopic laws as laws of large numbers}
Take a macroscopic box $\Lambda=[0,L]^d$ that  contain $N$ freely moving distinguishable particles and fixed obstacles of arbitrary shapes.  $N$ should be thought to be of the order of magnitude of the Avogadro number : $6\times 10^{23}$.   

\begin{figure}[h!]
\begin{center}
\includegraphics[width = .25\textwidth]{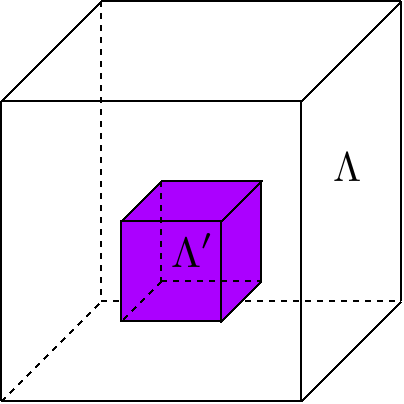} 
\end{center}
\label{LambdaPrime}
\caption{In the first experiment, the cloud of particles is initially concentrated in a volume $\Lambda'$.}
\end{figure}

\noindent A first experiment is performed on this box.  The $N$ particles are initially located in a cube $\Lambda'\subset\Lambda$ of side length $L'<L$, see Figure \ref{LambdaPrime}.
  The evolution of the density of the cloud of particles is monitored through a beam of light that crosses the system.  We call this density $\rho: \Lambda\times [0,\infty[\to \rho(\x,t)\in \bbR^+$. The initial state is described by $\rho(\x,0)={\bf 1}_{\Lambda'}(\x)/|\Lambda'|$.
The empirical fact that is observed  at the macroscopic level is that the  density evolves according to the laws of diffusion:
\begin{equation}
\left\{
\begin{array}{l}
\partial_t\rho(\x,t)=\kappa\,\Delta\rho(\x,t)\\
n_{\x}\cdot\nabla\rho(\x,t)=0, \quad \x\in \partial\Lambda\\
\rho(\x,0)=\rho_0(\x):={\bf 1}_{\Lambda'}(\x)/|\Lambda'|,
\end{array}
\right.
\label{diffusion}
\end{equation}
where  $n_{\x}$ is the vector normal to the  boundary of the box $\partial\Lambda$ at $\x$ and $\kappa$ is a strictly positive constant.  How can we explain this phenomenon from the motion of the individual atoms ?
For each {\it macroscopic} coordinate $\mathbf{x}=(x_1,\ldots,x_d)\in\Lambda$, we define a {\it microscopic} coordinate $\q=\mathbf{r}/\epsilon_N$ where  $\epsilon_N=\frac {1}{N^{1/d}}$. The motion of the particles is entirely determined by the law of Newtonian mechanics.  When a particle makes a collision with one of the fixed obstacles or the boundaries of the boxes, its velocity is modified according to the laws of specular reflection.  We assume that each particle starts with a speed equal to $1$.  Since this property is preserved by the dynamics, the microscopic motion of a given particle (with label $i\in\{1,\ldots, N\}$) is described by a map $t\to (\q_i(\epsilon^{-2}_Nt),\p_i(\epsilon^{-2}_Nt))\in [0,L/\epsilon_N]^d\times S^{d-1}$ where $S^{d-1}$ is the unit sphere in $d$ dimensions. The  coordinates $(\q_i,\p_i)$ are the {\it microscopic} positions and velocities of the $i$-th particle. The microscopic time-scale is $\epsilon^{-2}_N t$.  The scaling of the time variable is a priori arbitrary but is fixed here by the fact that the solution of the diffusion equation $\rho(\x,t)$ is invariant under the transformation $(\x,t)\to (\lambda \x,\lambda^2 t)$, $\lambda>0$. 

In the absence of any other information, we assume that the initial positions and velocities of the particles are independent and identically  uniformly distributed, with density
 \begin{equation}
 f(\q_i,\p_i)=\frac{\epsilon^d_N}{|\Lambda'||S^{d-1}|}{\bf 1}_{\Lambda'}(\epsilon_N \q_i){\bf 1}_{S^{d-1}}(\p_i)
 \label{uniform}
 \end{equation}
 for every $i=1,\ldots, N$.
  If the position of each particle is chosen independently of the others with that density, the number of particles in a microscopic volume of size of order $1$  follows a Poisson distribution with finite mean as $N\to\infty$.  We denote by $\bbP$ the law of probability of the initial positions and velocities of particles. No information about the spatial location or the shape of the obstacles in $\Lambda$ is known either. We denote by $\bbQ$ the probability distribution on those degrees of freedom. It is chosen such that in the limit $N\to\infty$, the number of obstacles in a microscopic volume of size $1$ follows a distribution with a finite mean and such that, almost surely, the dynamics of moving particles is well-defined at all time.   The dynamical system defined in this way is an instance of the {\it random Lorentz gas}.

We define  the empirical density of particles :
\begin{equation}
\rho_N(\x,t)=\frac 1 N\sum_{j=1} ^N\delta (\epsilon_N \,\q_j(\epsilon_N^{-2} t )-  \x )
\label{emp_density}
\end{equation}
The density $\rho_N$ contain all possible information about the density.  Indeed, it is easy to see that
$$
\rho_N(V,t):=\int_{\bbR^d}d\x\; {\bf 1}_V(\x)\rho_N(\x,t)
$$
gives the proportion of particles that belong to  any $V\subset \Lambda$ at time $t$.
It is straightforward to see that the following statement holds : if $\{(\q_i(0),\p_i(0)):i=1,\ldots,N\}$ is a collection of i.i.d variables with marginals given by (\ref{uniform}) then for any bounded  function $h$ and any $\delta>0$ :
\begin{equation}
\lim_{N\to\infty}\bbP[|\left<\hat\rho_N(0),h\right>-\left<\rho_0,h\right>|>\delta]=0,
\label{large_number0}
\end{equation}
where $\rho_0$ is given by (\ref{diffusion}), $\left<h,g\right>=\int_{\bbR^d}d\x\; h(\x)g(\x)$ and we use the notation $\hat\rho_N(t):=\rho_N(\cdot,t)$.
The goal is  to show the
\begin{conjecture}
There exists a natural \footnote{By natural we mean as uniform as possible over the locations and shapes of obstacles.} distribution $\bbQ$ such that for any $t>0$, any bounded function $h$ and any $\delta>0$ :
\begin{equation}
\lim_{N\to\infty}\bbP\times \bbQ[|\left<\hat\rho_N(t),h\right>-\left<\rho(t),h\right>|>\delta]=0.
\label{goal1}
\end{equation}
\label{large}
where $\rho(t):=\rho(\cdot,t)$ is the solution of (\ref{diffusion}) for some $\kappa>0$.
\end{conjecture}
The law of ordinary diffusion is therefore understood as a {\it law of large numbers}:  as $N$ becomes very large, the probability that the empirical density $\hat\rho_N(t)$ differs significantly from the solution of the diffusion equation goes to zero.

It is natural to consider first a simpler version of the problem in which the randomness of the obstacles is removed, i.e. $\bbQ$ is taken to be a Dirac distribution $\delta_C$ on a special configuration of obstacles giving rise to a chaotic dynamics.  This  is exactly  the result of Bunimovich and Sinai \cite{bunisinai}.  They consider the 2D case in which obstacles are disks located at the vertices of a regular lattice such that the induced billiard dynamics has a finite horizon \footnote{For instance, the center of each (sufficiently large) disk is located at a vertex of a triangular lattice.}.  Their result implies that for any $t>0$, any bounded function $h$ and any $\delta>0$ :
\begin{equation}
\lim_{N\to\infty}\bbP\times {\bf \delta}_C[|\left<\hat\rho_N(t),h\right>-\left<\rho(t),h\right>|>\delta]=0.
\label{bunisinai}
\end{equation}

Let us sketch how this statement is obtained. Let $h:\bbR^d\to\bbR$ a bounded  function.
First, one computes : 
\begin{eqnarray}
\left<\hat\rho_N(t),h\right>&=&\frac 1 N \int_{\bbR^d}\, d\x \sum_{j=1}^N\delta(\epsilon_N \q_i(\epsilon_N^{-2} t)-\x)h(\x)\\
&=&\frac 1 N\sum_{j=1}^Nh(\epsilon_N \q_j(\epsilon_N^{-2} t)).
\end{eqnarray}
Thus, because the initial positions of the particles are identically distributed :
\begin{eqnarray*}
\bbE[\left<\hat\rho_N(t),h\right>]&=&\bbE[h(\epsilon_N\q_1(\epsilon_N^{-2} t))].
\end{eqnarray*}
Next, the theorem 2 of \cite{bunisinai} implies \footnote{To be more precise Bunimovich and Sina\"i consider the case $L=\infty$ but their method should apply directly to the finite $L$ case} that
\begin{equation}
\lim_{N\to\infty}\bbE[h(\epsilon_N\q_1(\epsilon_N^{-2} t))]=\int_{\bbR^d}\rho(\x,t) h(\x,t)\,d\x
\label{chaos}
\end{equation}
where $\rho(\x,t)$ is the solution of (\ref{diffusion}).
To derive (\ref{chaos}), one has to rely on the strong chaotic properties of the billard system under study.

Next, since $\{\q_j(t): 1\leq j\leq N\}$ are {\it independent}, the variance of $\left<\hat\rho_N(t),h\right>$ is 
$$
{\mathrm {Var}}[\left<\hat\rho_N(t),h\right>]=\frac 1 N{\mathrm {Var}}[h(\epsilon_N\q_1(\epsilon_N^{-2} t))]=O(\frac 1 N)
$$
since $h$ is bounded.  Thus, Chebychev inequality allows us to conclude the proof of Conjecture \ref{large} in the case where $\bbQ=\delta_C$. One should note that the proof is made of two steps of very different levels of complexity.  The first step is basically given by (\ref{chaos}) and this is where the whole difficulty is located.  The second step is a concentration result of the random variable $\left<\hat\rho_N(t),h\right>$ around the expected value $\bbE[\left<\hat\rho_N(t),h\right>]$.  This part  is trivial because when $\bbQ$ is replaced by $\delta_C$, the positions and velocities of the particles remain independent for all time $t$.  The fact that it is so trivial is probably the reason why it is hard to find a reference where this step is mentioned or even alluded to.  It is however essential and when $\bbQ\neq \delta_C$, the statistical independence of the motions of particles is lost.  Controlling the correlations between them to ensure the concentration of $\left<\hat\rho_N(t),h\right>$ around its mean does require some work. We will see below that this issue arises in the mirrors model and how it can be dealt with.

\begin{figure}[thb]
\begin{center}
\includegraphics[width = .45\textwidth]{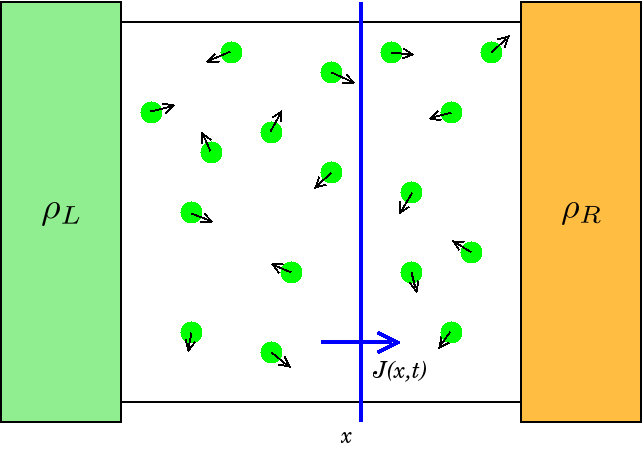} 
\end{center}
\label{Courant}
\caption{At the two sides of the cube $\Lambda$, particles reservoirs maintain constant values of the local densities of particles $\rho_L$ and $\rho_R$.}
\end{figure}
A second experiment may be performed on the box.  At the two sides of the cube $\Lambda$ perpendicular to $e_1=(1,0\ldots,0)$, particles reservoirs maintain constant values of the local densities of particles $\rho_L$ and $\rho_R$, respectively on the left and right side, see Figure 2.  

A device records the net flux of mass crossing a section of $\Lambda$ perpendicular to $e_1$ per unit time. This quantity is denoted by $j(x,t)$ when the section contains the point  $(x,0,\ldots,0)$ for  $x\in [0,L]$. After some transient time proportional to $L^2$, it is observed that the instantaneous current of particles takes the stationary value :
\begin{equation}
j_s(x)=\frac{\kappa}{L} (\rho_L-\rho_R).
\label{ficklaw}
\end{equation}

With respect to the first experiment, the coupling to external reservoirs introduce an additional probabilistic element.  The law of the reservoirs and the initial conditions of the particles inside the system is denoted by $\bbP$.
One can introduce an empirical current of particles per unit time $\hat J_N(t)$ \footnote{This quantity will be our main object of study in the next section and will be given a precise definition there.}.  Again, the goal is to show the following conjecture :
\begin{conjecture}
There exists a natural distribution $\bbQ$ such that for any bounded continuous function $h$ and any $\delta>0$ :
\begin{equation}
\lim_{N\to\infty}\lim_{t\to\infty}\bbP\times \bbQ[|\left<\hat J_N(t),h\right>-\left<j_s,h\right>|>\delta]=0.
\label{goal2}
\end{equation}
\label{large2}
\end{conjecture}
While this has never been done explicitly, one should expect that with the choice $\bbQ=\delta_C$ the result  follows from the methods of \cite{bunisinai}. In \cite{Basile}, the authors show that in a low density regime limit, the expectation of the stationary current (with respect to $\bbP\times \bbQ$ ) converges to $j_s$. 
The statement corresponding to Conjecture \ref{large2} together with the exponential convergence to the stationary current has been obtained in the case of a discrete space-time dynamics in \cite{Lefevere2}. 
We now outline how the problem may be tackled in the context of the mirrors model.
\section{The mirrors model}
The mirrors model was introduced by Ruijgrok and Cohen \cite{Ruijgrok} as a lattice version of the random Lorentz gas or the Ehrenfest wind-tree model.  The latter encompasses a Iarge class of models in which obstacles do not induce a chaotic behaviour of the trajectories of the particles.  A  fundamental question which remains open regarding those models is  whether a non-chaotic deterministic dynamics may give rise to a {\it macroscopic} diffusive behaviour.
In the mirrors model, particles travel on the edges of the cubic lattice generated by $\bbZ^d$.  ``Mirrors" are located at the vertices of the lattice and deflect the motion of an incoming particle in a new direction, see Figure \ref{mirrors_model} for an illustration in the 2D version of the model.  The precise general definition is given below.
\begin{figure}[thb]
\begin{center}
\includegraphics[width = .40\textwidth]{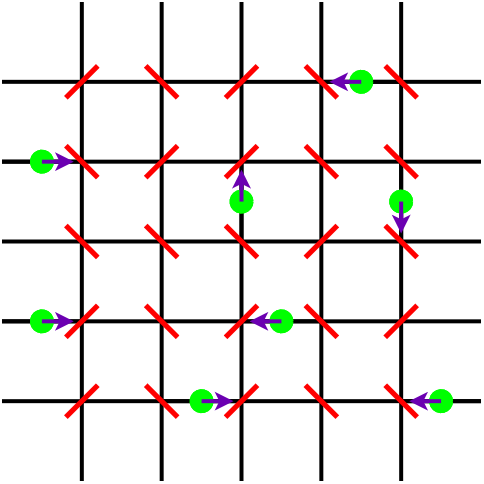} 
\end{center}
\caption{In the mirrors model on $\bbZ^2$, green particles travel on the edge of the lattice and are deflected by mirrors located at the vertices. The green particles do not interact with each other. }
\label{mirrors_model}
\end{figure}

A quick look at the structure of the orbits reveals its total lack of ergodicity.  Indeed, in Figure \ref{crossing_orbits} a sample of orbits in a finite box with periodic boundary conditions in the vertical direction and reflecting boundary conditions in the horizontal direction is pictured with different colors. For almost any configuration of the mirrors, no orbit is able to visit the entire phase space.   

 A perhaps even more striking  fact is that, in any dimension, the motion of a particle in an environment of randomly orientated mirrors is not  a gaussian diffusion \footnote{ In spite of this, it has been observed numerically \cite{Kong} that in two dimensions, the mean-square displacement of a given particle is linear in time, allowing the definition of a {\it microscopic} diffusion coefficient.  This is of course a much weaker property than the property (\ref{goal2}).  In particular we will see that one can not define a {\it macroscopic} diffusion coefficient in $2D$. } \cite{BT}. More precisely this means that (\ref{chaos})  (where the expectation is taken with respect to $\bbP$ and $\bbQ$) does not hold. 
 
 Our goal is to show that in spite of these unpromising properties,  the mirrors model does exhibit normal macroscopic conductive properties when $d\geq 3$ in the sense that the analogue of  (\ref{goal2}) holds.     It turns out that quite weak conditions on the statistics of orbits are sufficient to ensure the validity of Fick's law at the macroscopic level.  It is therefore not necessary that orbits behave as a Gaussian diffusion to ensure the validity of Fick's law.  Thus, the normal {\it macroscopic} laws of diffusion apply to a much wider class of dynamical systems than generally expected.  

The dynamics of the mirrors model is reversible in the usual sense of the word in the context of Hamiltonian dynamics. Namely, under the reversal of the velocities of all particles at a given time $t>0$, the dynamics brings the system of particles to its initial condition at time $0$ (with reversed velocities), see (\ref{reversible}).  This reversibility property of the dynamics will allow us to show that when the system is large, the number of orbits travelling from one side of the system to the other one basically determines the value of the current in the stationary state.  This will allow to formulate a condition on the distribution of orbits that is both sufficient and necessary for the validity of Fick's law.

We recall now briefly the set-up of the original mirrors model. Particles travel on the edges of  $\bbZ^2$ with unit speed.
Mirrors are located at some vertices of the lattice and take two possible angular orientations : $\{\frac \pi 4, \frac{3\pi}{4}\}$.
 When a particle hits a mirror, it gets deflected according to the laws of specular reflection, see Figure \ref{crossing_orbits} for sample trajectories of particles. It is convenient to think that every particle starts at time zero with a given velocity at a vertex of the lattice $\cQ$ that is obtained by taking the middle point of every edge of $\bbZ^2$. As all particles move with unit velocity, one can simply  observe the evolution of the system at discrete times $t\in\bbN$.  At those times, the particles will be always located at one of the vertices of the new lattice $\cQ$ with a well-defined velocity.  In general, the orientation of the mirrors is picked randomly.  It is obvious that the motion of a single particle can not be described as a Markov process.  When a particle hits a mirror for the second time, no matter how far back in the past the first visit occurred, its reflection is strongly affected by the way its was reflected at the first visit. For instance in Figure \ref{crossing_orbits}, the two orientations of the mirrors are picked at random, and in that case, at the second visit the reflection is always deterministic.
 
\begin{figure}[thb]
\begin{center}
\includegraphics[width = .40\textwidth]{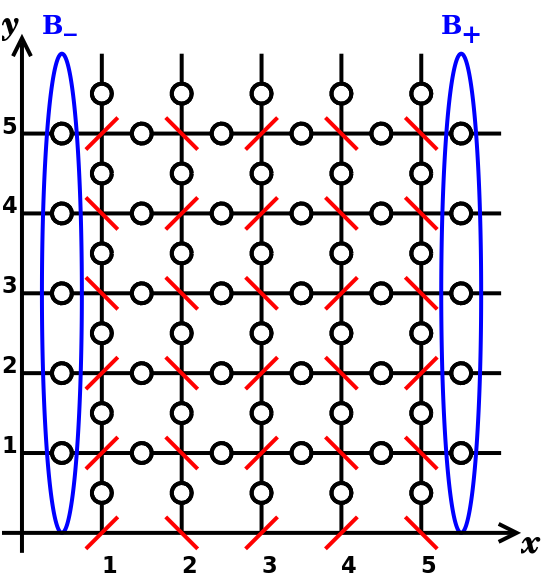} 
\end{center}
\caption{The spatial component of the phase space $\cM$  and of the sets $B_-$ and $B_+$ in $2D$.}
\label{mirrors1}
\end{figure}

\begin{figure}[thb]
\begin{center}
\includegraphics[width = .40\textwidth]{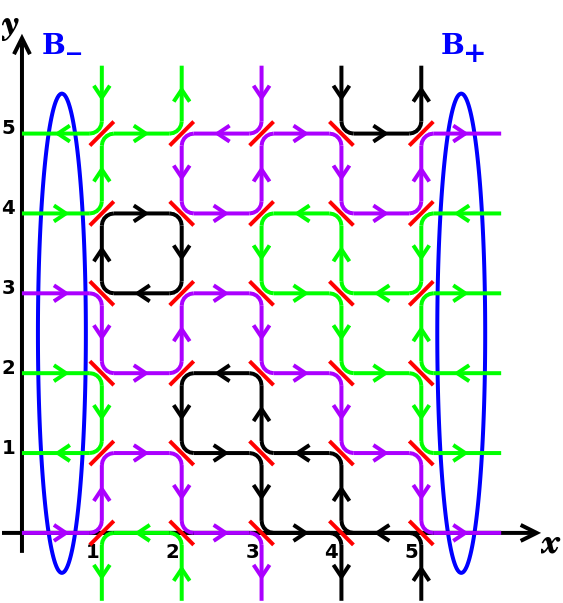} 
\end{center}
\caption{$N=6$. Crossing orbits are coloured in purple, internal loops in black and non-crossing orbits are coloured in green. The travel direction given by the arrows is arbitrary.  Each edge of the crossing orbits will be used twice in a given orbit : once in each direction. For this configuration of mirrors $\cN=2$.}
\label{crossing_orbits}
\end{figure}

We come now to a more general definition of the dynamics in $d$ dimensions.  We denote by $\z=(z_1,\ldots,z_d)$ a generic element of $\bbZ^d$.  As for $\bbZ^2$, we consider the set of midpoints of edges of an hypercube of $\bbZ^d$  of side $N$ and with periodic conditions in all but the first direction.  We call this set $\cQ$.  It may be described as follows : $\cQ=\bigcup_{i=1}^d L_i$
where $L_i=\left\{\z+\frac 1 2 \e_i: \;  0\leq z_1 \leq N-1,\;(z_2,\ldots,z_{d})\in (\bbZ/N\bbZ)^{d-1}\right\}$.
Let $(\e_1,\ldots,\e_d)$ the canonical basis of $\bbR^d$, the space of possible velocities is $\cP=\{\pm\frac{\e_1}{2},\ldots,\pm\frac{\e_d}{2}\}$ and
the phase space of the dynamics is 
$$
\cM=\{(\q,\p): \q\in\cQ,\p\in\cP\;{\rm s.\; t.\; if}\;\q \in L_i\;{\rm then}\; \p=\pm\frac{\e_i}{2}\}.
$$
 We denote a generic point of $\cM$ by $(\q,\p)$. The set of points in $\cM$ whose spatial coordinate belongs to the boundaries of the system is  $B=B_-\cup B_+$, with
\begin{eqnarray}
B_-&=&\{x=(\q,\p)\in\cM: \q=(q_1,\ldots,q_d) \in L_1, q_1=\frac 1 2\}\nonumber\\
B_+&=&\{x=(\q,\p)\in\cM: \q=(q_1,\ldots,q_d)\in L_1, q_1=N-\frac 1 2\}.\nonumber
\end{eqnarray}
See Figure \ref{mirrors1}.

For each $\z\in\bbZ^d$, we define the action of a ``mirror" on the velocity of an incoming particle by $\pi(\z;\cdot)$ which is a bijection of $\cP$ into itself. It satisfies the following conditions :
\begin{equation}
\left\{
\begin{array}{l}
\pi(\z;-\pi(\z;\p))=-\p,\quad \forall \z\in\bbZ^d,\forall\p\in\cP\\
\pi(0,z_2,\ldots,z_{d};- \frac{\e_1}{2})= \frac{\e_1}{2}\\
\pi(N, z_2,\ldots,z_{d}; \frac{\e_1}{2})=- \frac{\e_1}{2}, \;(z_2,\ldots,z_{d})\in (\bbZ/N\bbZ)^{d-1}
\end{array}
\right.
\label{conditionsx}
\end{equation}
The dynamics is defined on $\cM$ in the following way.   For any $(\q,\p)\in\cM$ :
\[
F(\q,\p)=\left(\q+\p+ \pi(\q+\p;\p),\pi(\q+\p;\p)\right).
\]
\noindent It is easy to check that the map $F$ is a bijection on $\cM$. The two last conditions in (\ref{conditionsx}) are just saying that when particles hit the boundaries they are reflected backwards.
 We define an operator $R:\cM\to\cM$  which reverses the velocities by 
$R (\q,\p)=(\q,-\p)$ and we see that the first of the conditions (\ref{conditionsx}) ensures that the map $F$ is {\it reversible},
\begin{equation}
F^{-1}=R F R.
\label{reversible}
\end{equation}

We define the orbit of a point $x\in\cM$, ${\mathcal O}_x=\{y\in\cM: \exists t\geq 0,F^t(x)=y\}$
and its period, $T(x)=\inf\{t\geq 0: F^t(x)=x\}$.  From the fact that $F$ is bijective, one infers that for every $x\in\cM$, ${\mathcal O}_x$ is a loop :  $T(x)\leq |\cM|$ and that orbits are non-intersecting : if $ y\notin {\mathcal O}_x$, then ${\mathcal O}_x\cap{\mathcal O}_y=\emptyset$. A given orbit is also non-self-intersecting : if $y\in {\mathcal O}_x$ and $y\neq x$ then $F(y)\neq F(x)$.

As we are interested in the transport of particles, we define occupation variables $\sigma(\q,\p;t)\in\{0,1\}$ that record the absence or presence of a particle at position $\q$ with velocity $\p$ at time $t\in\bbN$.
When connecting the system to external particles reservoirs, we obtain the following evolution rule : given $\sigma(\cdot;t-1)$, we define $\sigma(\cdot;t)$ for all $t\in\bbN^*$ by
$$
\sigma(x;t)=\left\{
\begin{array}{lll}
\sigma(F^{-1}(x);t-1)\quad {\rm if} \quad x\notin B_{-}\cup B_{+}\\
\sigma^-_{x}(t-1)\quad {\rm if}  \quad x\in B_{-}\\
\sigma^+_{x}(t-1)\quad {\rm if} \quad  x \in B_{+}.
\end{array}
\right.
$$
The families of random variables $\{\sigma^-_{x}(t):x\in B_{-},\, t\in\bbN\}$ and $\{\sigma^+_{x}(t): x\in B_{+},\, t\in\bbN\}$  consist of independent Bernoulli variables with respective parameters $\rho_-$ and $\rho_+$.
If one chooses $\{\sigma(x;0):x\in\cM\}$ to be a collection of independent  random variables, then it is easy to see by induction that at any $t\geq 0$,  $\{\sigma(x;t):x\in\cM\}$ is a collection of i.i.d Bernoulli random variables. To simplify a bit the discussion, we choose an homogeneous initial distribution, i.e. all Bernoulli random variables have a common parameter $\rho_I$.  The distribution of the collection $\{\sigma(x;t):x\in\cM\}$ becomes stationary after a {\it finite} time.  More precisely,  for any $t\geq |\cM|$, we have the following equality in law :
$$
\sigma(x,t)=\left\{
\begin{array}{lll}
\sigma_I\quad {\rm if} \quad  \cO_x\cap B=\emptyset\\
\sigma_-\quad {\rm if} \quad F^{-t^*}(x)\in B_-\\
\sigma_+\quad {\rm if} \quad F^{-t^*}(x)\in B_+
\end{array}
\right.
$$
where $t^*=\inf\{t: F^{-t}(x)\in B\}$ and $\sigma_\pm$ and $\sigma_I$ are Bernoulli random variables of parameter $\rho_{\pm}$ and $\rho_I$.

 Proceeding  as in \cite{Lefevere2}, it is possible to show that when the size of the system goes to infinity, the stationary current converges in probability to the proportion of crossing orbits times the chemical potentials difference.  We define the average current of particles that crosses the hyperplane $\cQ^l=\{\q\in\cQ: q_1=l+\frac 1 2\}$, $l\in\{1,\ldots,N-2\}$ during a diffusive time interval $N^2$  :
\begin{equation}
J(l,t)=\frac 1 {N^{d+1}}\sum_{s=t+1}^{t+N^2}\sum_{x\in\cM}\sigma(x;s)\Delta(x,l)
\label{cud}
\end{equation}
where $\Delta(x,l)=2(\p\cdot\e_1){\bf 1}_{\q\in \cQ^l}$,with $x=(\q,\p)$.
Thus $\Delta(x,l)$ takes the value $+1$ (resp. $-1$) if  $x$ crosses the slice $\cQ^l$ from left to right (resp. from right to left). We denote by $\cN_{\pm}$ the numbers of crossings from $B_{\pm}$ to $B_{\mp}$ induced by $F$, i.e. $\cN_{\pm}=|S_\pm|$ where $S_\pm$ is given by
\begin{eqnarray}
S_\pm=\{ x\in B_{\pm}:\exists s>0,\forall 0<j<s,\; F^j(x)\notin B_{\pm}, F^s(x)\in B_{\mp}\}.\nonumber
\end{eqnarray}

One notes that $\cN_+=\cN_-$.  Indeed, since every orbit is closed, it must contain as many left-to-right than right-to-left crossings.  Thus, we set $\cN= \cN_+=\cN_-$.  Proceeding as in \cite{Lefevere2},  we get that for any $t\geq |\cM|$, $\bbE[J(l,t)]=\frac{\cN}{N^{d-1}}(\rho_--\rho_+)$.
\footnote{This relation implies that the average current flows in the ``right" direction and that when $\rho_-\neq\rho_+$, the average current in the stationary state is different from $0$ if and only if $\cN\neq 0$.}
Moreover, for every $\delta>0$, any $t\geq |\cM|$ and  $l\in\{1,\ldots,N-2\}$,

\begin{equation}
\bbP\left[\left|J(l,t)-\frac{{\mathcal N}}{N^{d-1}}(\rho_--\rho_+)\right|\geq \delta \right]\leq 2\exp(-\delta^2N^{d+1}).
\label{JN}
\end{equation}

We take now random configurations of reflectors $\{\pi(\z;\cdot): \z \in \bbZ^d\}$.  The law of the reflectors is denoted by $\bbQ$.  The map $F$ becomes now a random map.

The model satisfies {\it Fick's law} if and only if  there exists some $\kappa>0$ (the conductivity) such that $\forall\delta>0$,
\begin{equation}
\lim_{N\to\infty}\lim_{t\to\infty}\bbP\times\bbQ[|N J(l,t)-\kappa(\rho_--\rho_+)|>\delta]=0.
\label{Ficks}
\end{equation}
As in \cite{Lefevere2},  it is easy to infer from (\ref{JN}) that the following theorem holds.
\begin{theorem}{\bf Sufficient and Necessary Condition for Fick's law} :
(\ref{Ficks}) holds if and only if there exists $\kappa>0$ such that for any $\delta>0$,
\begin{equation}
\lim_{N\to\infty}\bbQ\left[\left|\frac{\cN}{N^{d-2}}-\kappa\right|>\delta\right]=0.
\label{Ficks2}
\end{equation}
\end{theorem}
We see that the central object to study is the distribution of the number of crossing orbits $\cN$.  The expectation of this quantity is related to the probability that one orbit crosses the system, while the variance is given in terms of the joint probability that orbits with two different starting points cross the system.  Indeed  by periodicity, we have, using the notations $O=((\frac 1 2,0,\ldots,0), \frac{\e_1}{2})$ and $S=S_-$ :
\begin{equation}
\bbE\left[\frac{\cN}{N^{d-2}}\right]=\frac N {N^{d-1}}\sum_{x\in B_{-}}\bbE[{\bf 1}_{x\in S}]=N\bbQ[O\in S]
\label{expectation}
\end{equation}
and
\begin{eqnarray}
{\rm Var }\left[\frac{\cN}{N^{d-2}}\right]
=\frac{1}{N^{2d-4}}\sum_{x,y\in B_{-}}\delta(x,y)=\frac{1}{N^{d-3}}\sum_{x\in B_{-}}\delta(O,x)\nonumber\\
\label{variance}
\end{eqnarray}
with 
\begin{equation}
\delta(x,y)=\bbQ[x\in S,y\in S]-\bbQ[x\in S]\bbQ[y\in S].
\label{corel}
\end{equation}

Thus if the two following 
\begin{Conditions}  are satisfied : 

\begin{enumerate}
\item There exist $\kappa>0$ such that the RHS of (\ref{expectation}) converges to $\kappa$ as $N\to\infty$.
\item  The RHS of (\ref{variance}) goes to zero as $N\to\infty$.
\end{enumerate}
\label{conditions}
\end{Conditions}
then Fick's law (\ref{Ficks}) holds in the stationary state.
We note first that when $d=2$, (\ref{Ficks2}) can not hold, whatever the distribution $\bbQ$ is.  To see this, we adapt an argument found in \cite{Kozma}.
  Indeed, the spatial part of each crossing orbit crosses any ``vertical" section $\cQ^l$ an odd number of times.  On the other hand, the spatial part of any non-crossing orbit must cross any vertical section an even number of times, see Figure \ref{crossing_orbits}.
Thus,  $N$ and $\cN$ must have the same parity.   This implies that there can not exist $\kappa>0$ such that (\ref{Ficks2}) holds when $d=2$.   The origin of this issue lies in the strong correlations between crossing orbits that are present in two dimensions.

We turn now to the higher dimensional case $d\geq 3$ equipped with some natural and spatially  homogeneous distribution $\bbQ$. Now observe that if $\bbQ[\pi(\z;\frac{e_i}{2})=-\frac{e_i}{2}]>0$ for some $i=1,\ldots,d$ then an orbit starting from $O$ will encounter this type of reflecting mirror after an exponential number of steps and therefore $\bbQ[0\in S]\leq e^{-cN}$ for some $c>0$.  This, in turn, implies that $\lim_{N\to\infty}N\bbQ[0\in S]=0$ and that Fick's law can not hold.  Thus from now on, we consider maps such that   $\pi(\z;\frac{e_i}{2})\neq-\frac{e_i}{2}$ if $0<z_1 < N$ and such that the conditions (\ref{conditionsx}) are satisfied. We call the set of such maps $\Pi$.  We take $\bbQ$ such that the  collection of maps
\begin{equation}
\{\pi(\z;.): 0<z_1 < N,\;(z_2,\ldots,z_{d})\in (\bbZ/N\bbZ)^{d-1}\}
\nonumber
\end{equation}
is independent and that each map is uniformly distributed  over $\Pi$. We note first that if the law of an orbit with respect to $\bbQ$ was similar to the law of a simple random walk, then there would be a $\kappa>0$ such that $\lim_{N\to\infty}N\bbQ[0\in S]=\kappa$, this follows from the gambler's ruin argument.  Similarly, if the orbits were independent objects, then the RHS of (\ref{variance})  would go to zero because the only non-zero term would be the one with $x=O$ and $\bbQ[O\in S]\sim\kappa/N$.  We also note that the average stationary current
is identified as the difference between chemical potentials times
the probability that a particle crosses the system, an idea that was put forward in \cite{Carlos}, in the context of chaotic systems. The law $\bbQ$ of the mirrors induces a law on the set of orbits which is a priori very far from the distribution of independent simple random walks.  The set of orbits is a very interesting lattice object in itself which features some (self-)avoiding properties as we mentioned above. 

\begin{figure}[h!]
\begin{center}
\includegraphics[width = .45\textwidth]{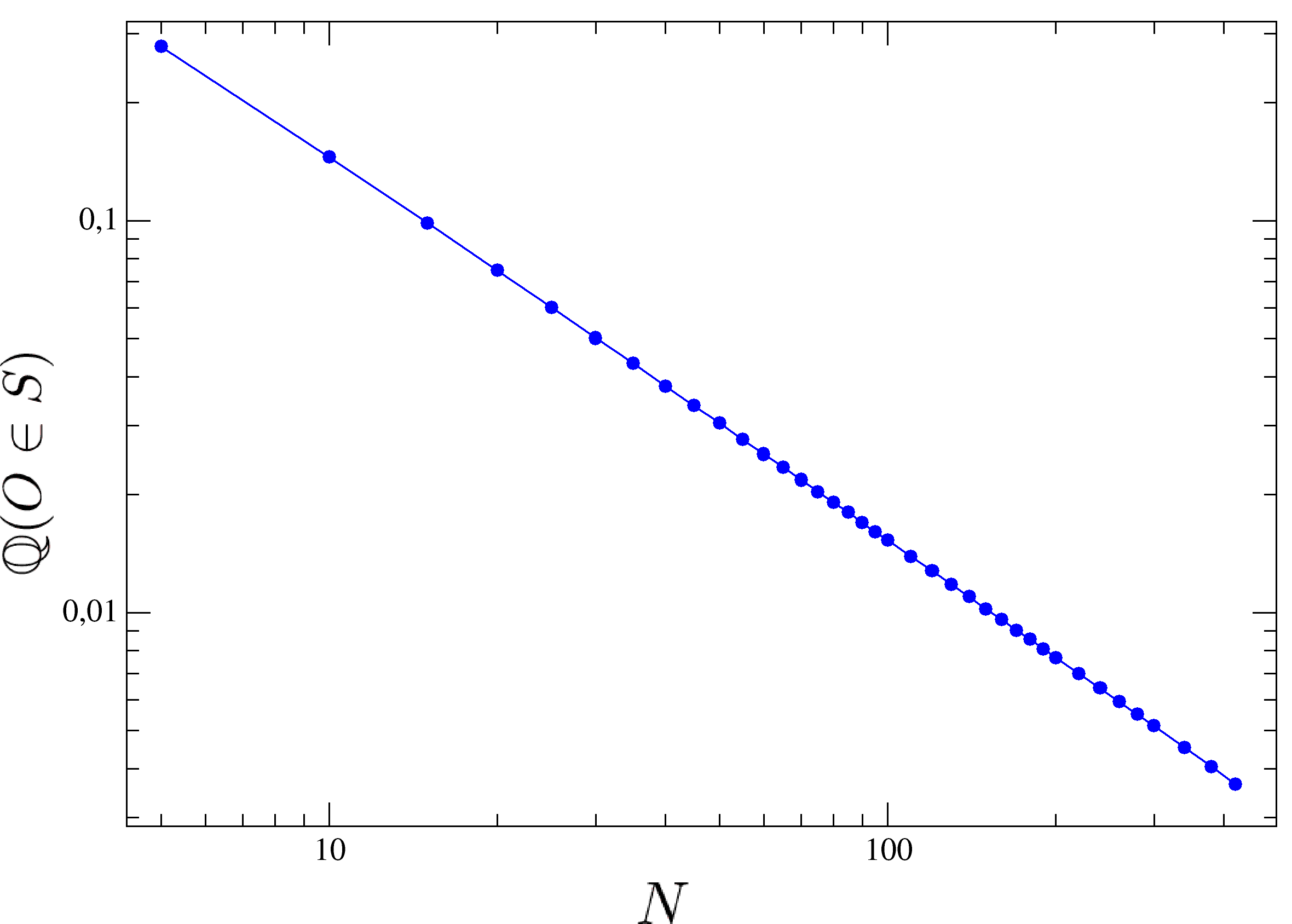} 
\end{center}
\caption{$\mathbb Q(O\in S)$ for $N$ from $5$ to $420$.  The 95\% confidence interval is about half the size of a dot.}
\label{3DPcrossing}
\end{figure}

Fortunately, what is needed to ensure the validity of (\ref{Ficks2}) is much less than the full joint distribution of the orbits.  Thanks to (\ref{expectation}) and (\ref{variance}), one only has to analyze the marginal of a path starting on the boundary and also the joint probability of two such paths.
The distribution of a path starting at $O$ (i.e. on the boundary) is similar to the  one of a ``true" self-avoiding random walk \cite{Amit} but defined on $\cQ$ rather than on $\bbZ^d$ and with further constraints.  The diffusive behaviour of those walks for $d\geq3$ has been conjectured in \cite{Amit}, see also the rigorous results of \cite{Toth}. It can be expected that as the dimensionality of the system increases, the effect of the revisits of an orbit to the same mirror decreases. In a process where the mirrors are flipped randomly after being used (i.e memory effects are killed), we computed that in $d=3$ the crossing probability is $\sim 3/2N$.
Numerical simulations in $d=3$ show that this number is indeed a good approximation. The log log plot of the crossing probability $\bbQ[O\in S]$ is given in Figure \ref{3DPcrossing} for $N$ up to $400$.  The corresponding conductivity is $\kappa=1.535\pm0.005$.    As the conductivity measured in simulations is slightly higher than $3/2$, it indicates that recollisions tend to push forward the orbit.  

We must show now that $\sum_{x\in B_{-}}\delta(O,x)\to 0$ as $N\to\infty$.  We know that this sum is positive because it is a variance, and thus it is enough to get an upper bound on the sum.  We will use a numerical analysis to show that $\delta(O,x)<0$ for all but a finite (i.e. independent of $N$) number $x\in B_-$. 

But before doing that and to get a better picture of the origin of the correlations $\delta(0,x)$, we split them in two parts. Given an orbit $\cO$, we denote by $\gamma(\cO)$ the set of edges of $\bbZ^d$ used by $\cO$.  For each $\z\in Z_N:=\{\z\in\bbZ^d: 1\leq z_1\leq N-1, (z_2,\ldots,z_d)\in (\bbZ/N\bbZ)^d\}$, let  also  $b_{\z}(\cO)$ be the half of the number of times that the orbit $\cO$ visits the vertex $\z$. Two {\it crossing orbits} $\cO$ and $\cO'$ are incompatible if $\gamma(\cO)\cap\gamma(\cO')\neq \emptyset$ and compatible otherwise. 
The law of a given {\it crossing} orbit is :
\begin{equation}
\bbQ(\cO)=\prod_{\z\in Z_N}\prod_{j=1}^{b_{\z}(\cO)}\frac{1}{2(d-j)+1}.
\label{law_single}
\end{equation}
If $\bbQ(\cO,\cO')$ is the joint probability of two orbits $\cO$ and $\cO'$ then $\bbQ(\cO,\cO')=0$ when $\cO$ and $\cO'$ are incompatible. If they are compatible, the joint law of two crossing orbits is given by
\begin{equation}
\bbQ(\cO,\cO')=\prod_{\z\in Z_N}\prod_{j=1}^{b_{\z}(\cO)+b_{\z}(\cO')}\frac{1}{2(d-j)+1}.
\label{law_joint}
\end{equation}
In particular, if they do not share any mirrors, then $\bbQ(\cO,\cO')=\bbQ(\cO)\bbQ(\cO')$.  
From those properties, starting from (\ref{corel}), we obtain that for $x\in B_-$,
\begin{eqnarray}
\delta(O,x)&=&\sum_{\cO_0,\cO_x} (\bbQ(\cO_0,\cO_x)-\bbQ(\cO_0)\bbQ(\cO_x))\nonumber\\
&-&{\sum_{\cO_0,\cO_x}}'\bbQ(\cO_0)\bbQ(\cO_x).
\label{decomp}
\end{eqnarray}
Both sums run over orbits that cross the box $\cQ$.   The first sum runs over compatible orbits such that $\gamma(\cO_0)$ and $\gamma(\cO_x)$ share a vertex of $\bbZ^d$.  The second (prime) sum runs over {\it incompatible} orbits $\cO_0,\cO_x$.   

Thus, from (\ref{decomp}), we see that correlations $\delta(O,x)$ are created from two opposite origins, corresponding to each of the two sums in (\ref{decomp}). If two orbits $\cO$ and $\cO'$  share some mirrors, then it is easy to see  from (\ref{law_joint}) that $\bbQ(\cO,\cO')>\bbQ(\cO)\bbQ(\cO')$. This in turn implies that the first sum is strictly positive.   This is a ``cooperative" effect, the orbits help each other crossing the system. The second sum corresponds to a {\it jamming} effect : an orbit starting from $O$ and crossing the system occupies a certain number of horizontal edges.  Because distinct orbits can not share the same edges, the occupied edges are no more available for an orbit starting from $x\in B_-$, this  creates negative correlations. 

\begin{figure}[h!]
\begin{center}
\includegraphics[width = .60\textwidth]{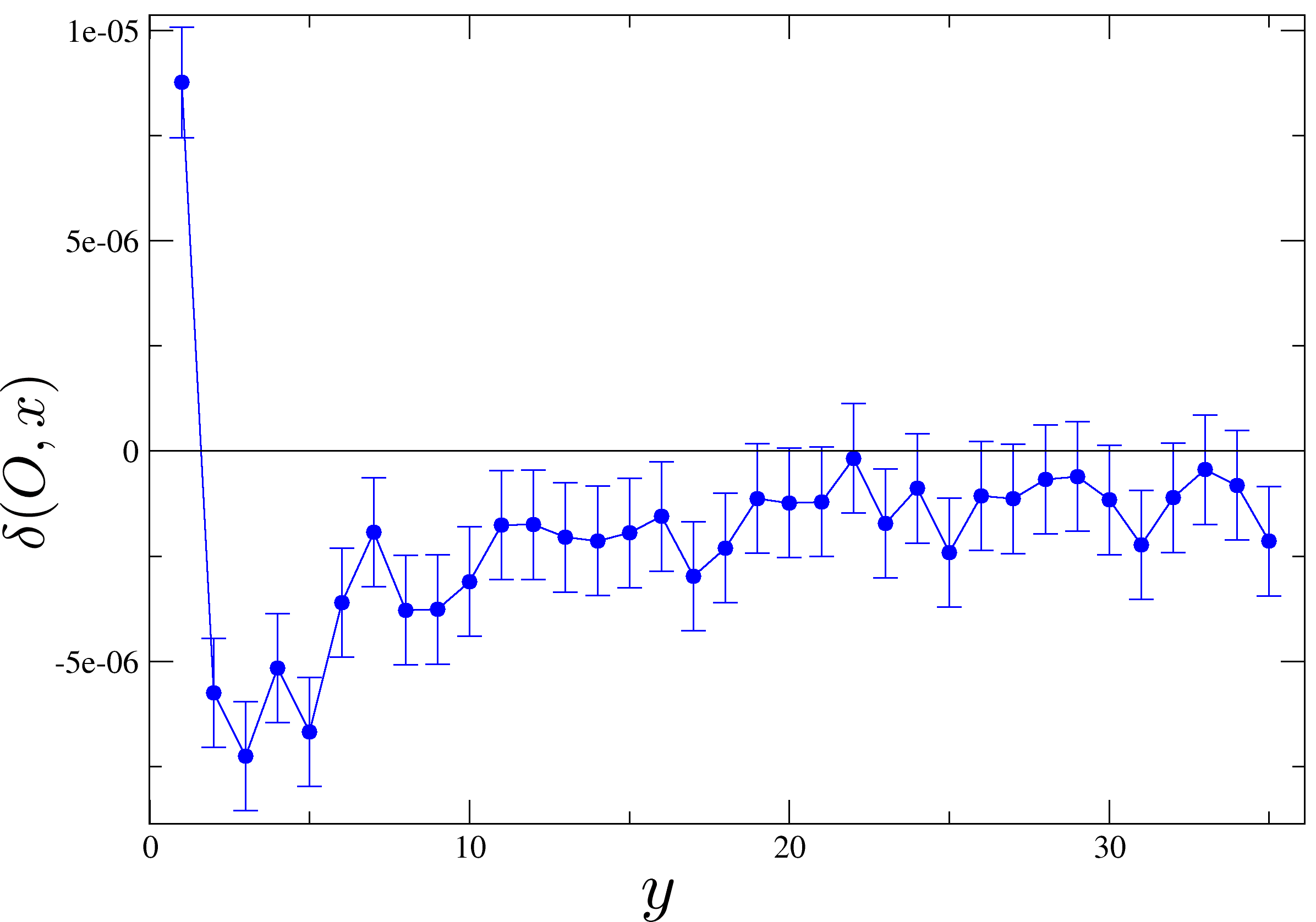} 
\end{center}
\caption {$\delta(O,x)$ for $x=((1/2,y,0),\frac{\e_1}{2})$. N=70. We draw the 95\% confidence interval.}
\label{3Dcorrelations}
\end{figure}

Numerical simulations in $d=3$ show that the latter effect dominates.   For all but a few points, the correlations $\delta(O,x)$ for $x\neq O$ are not only small but negative within confidence intervals, see Figure \ref{3Dcorrelations}.  The only exceptions are points $((1/2,1,0),\frac{\e_1}{2})$, $((1/2,0,1),\frac{\e_1}{2})$, $((1/2,N-1,0),\frac{\e_1}{2})$ and $((1/2,0,N-1),\frac{\e_1}{2})$ which give clearly positive correlations.  However, we checked that for $N=70$, $\sum_{y=1}^{N-1}\delta(O,((1/2,y,0),\frac{\e_1}{2}))=-1.360\times10^{-04}\pm 1.47\times10^{-05}$, i.e. it is negative with a margin of more than $9\sigma$.  $\sum_{z=1}^{N-1}\delta(O,((1/2,0,z),\frac{\e_1}{2}))$ must be equal by symmetry. Increasing values of $N$ do not modify this behaviour.  In particular, the number of points with positive correlations do not increase.  Since we know already that $\bbQ[O\in S]\sim\kappa/N$, as $N\to\infty$, we conclude with the same margin that $\sum_{x\in B_{-}}\delta(O,x)\leq \kappa/N\to 0$, as $N\to\infty$.

We expect the same behaviour in $d>3$.
A rigorous proof that the {\it crossing conditions} introduced above  are satisfied seems to be within reach in the present model.
Moreover, it is possible to draw a general conclusion from the above discussion.  If one is really interested in deriving macroscopic laws from microscopic dynamics, many detailed properties  of the latter are irrelevant.  Only weaker properties than chaoticity, ergodicity or Gaussian behaviour of the orbits are required.  In  the present context, the minimal properties necessary to obtain Fick's law are encapsulated in the crossing conditions.  It is of course natural to seek similar weak conditions in different contexts as for instance in the problem of the derivation of Fourier's law.


\end{document}